\documentclass{article}
\usepackage[utf8]{inputenc}
\usepackage{amsmath}
\usepackage{graphicx}
\usepackage{hyperref}
\usepackage[left=2cm,right=2cm,
    top=2cm,bottom=3cm,bindingoffset=0cm]{geometry}

\title{Computer Modelling of Bioheat Transfer for the Analysis of Brightness Temperature Distributions}

\author{Maxim V. Polyakov and Illarion E. Popov}

\date{\textit{Volgograd State University, Universitetsky pr., 100, Volgograd 400062, Russia}}

\begin{document}

\maketitle

\begin{abstract}
This paper presents a comprehensive computer simulation of thermal processes in multilayered biological tissues for the analysis of luminance temperature distributions recorded by microwave radiometry. A mathematical model combining the bioheat transfer equation with the electrodynamic description of electromagnetic field propagation in an inhomogeneous medium has been developed. The model accounts for variations in the thermophysical and dielectric properties of tissues, as well as convective heat exchange with the environment. A parametric analysis was performed to investigate the effects of ambient temperature, heat transfer coefficient, and tissue thermal conductivity on brightness temperature formation. Computational experiments revealed a clear linear dependence of the radiometric signal on both external and internal parameters. It was found that an increase in air temperature by 8 $^\circ$C causes a shift in the brightness temperature by about 2 $^\circ$C, which exceeds the typical error of the microwave radiometry method and emphasises the need for strict temperature control of the measurement conditions. The developed software system was validated against analytical solutions and experimental data, confirming the accuracy of the calculated temperature fields to within 0.1 $^\circ$C. The obtained results lay the groundwork for refining heat transfer models in biological tissues, improving the accuracy of non-invasive thermodiagnostic methods, and integrating computer modelling with machine learning algorithms to enable the automated interpretation of radiometric data. 

\vspace{2mm}
\end{abstract}

\section{Introduction}

The analysis of temperature distribution in biological tissues is of great importance in medicine and bioengineering, as it enables the detection and monitoring of pathological conditions such as inflammatory processes, tumour formation, and circulatory disorders \cite{Gautherie1982}. Changes in the temperature profile of tissues often serve as early markers of disease, making the study of thermal processes an important tool for diagnosis and evaluation of therapeutic \mbox{efficacy \cite{Zhen2015}.} One of the promising methods of non-invasive temperature monitoring is microwave radiometry (MWR), which is based on the measurement of deep (brightness) tissue temperature without physical influence \cite{Bardati2008, Vesnin2018}. However, the accuracy of this method is influenced by various factors, such as environmental conditions and the thermophysical properties of biological tissues, particularly their thermal conductivity. Detailed mathematical modelling of thermal processes is necessary to correctly interpret experimental data and improve the reliability of measurements.

Existing approaches to modelling thermal processes in biotissues include analytical solutions of heat conduction equations, numerical methods, and experimental studies \cite{Haemmerich2022, Akulova2024}. In particular, mathematical modelling based on the Pennes equation \cite{Pennes1948} and its modifications is widely used to evaluate the influence of various parameters on temperature profiles \cite{Avila2017, Figueiredo2019}. However, issues related to the effects of changing external conditions, the dynamics of heat exchange between the tissue and its environment, and the impact of tissue heterogeneity on temperature distribution remain unresolved. These aspects are critical for improving the accuracy of microwave radiometry and for the interpretation of the data obtained.

In the early stages of development, malignant tumours are characterised by increased metabolic activity, which leads to a local increase in heat \cite{Gautherie1982}. This effect is due to intensification of angiogenesis, changes in cellular respiration and other biochemical rearrangements accompanying tumour growth. Differences in the temperature profiles of healthy and diseased tissues formed the basis for non-invasive diagnosis using microwave radiometry, which measures deep thermal fields without physical intervention. Pioneering work in this field, including the \cite{Barett1980} study, has confirmed the clinical potential of the technique to detect malignancies. The key advantages of radiometry are safety, lack of radiation exposure and the possibility of multiple monitoring, which is particularly important for dynamic follow-up of patients \cite{Sedankin2018}. In recent years, the development of the method has focused on improving its diagnostic accuracy.
Current research, such as \cite{Zamechnik2019, Levshinskii2020}, focuses on improving instrumentation to enhance measurement sensitivity, developing signal processing algorithms to increase specificity, and integrating radiometry with other imaging modalities (e.g. thermography or MRI). Particular successes have been achieved in breast cancer screening, where the combination of high resolution and safety makes the method a promising alternative to traditional approaches \cite{Sedankin2018}. Simulation results are actively applied in the design of new medical antennas and radiometers with improved technical and performance characteristics, which contributes to the accuracy and reliability of diagnostic measurements \cite{Sedankin2018}. In particular, the ongoing research allows us to optimise the antenna design to improve sensitivity to deep temperature anomalies in biological tissues.

Current research in the field of computer modelling of thermal processes in biological tissues demonstrates significant progress in the development of mathematical approaches for the analysis of temperature distributions. Of particular interest are developments that take into account relaxation effects of heat transfer, where an analytical solution using the \cite{Ragab2021} equation for spherical biological structures has been proposed. An important direction is the consideration of nonlinear effects and complex multilayer structure of biological tissues. The developed finite element method for the nonlinear hyperbolic biothermia equation allowed us to take into account the temperature dependence of key parameters such as blood perfusion and metabolic heat generation \cite{Marin2021}. Further development of this approach demonstrated the possibility of determining the localisation of tumour masses through the analysis of temperature distribution in a five-layer skin model \cite{Akulova2024}. A significant part of research is devoted to applied aspects of modelling, especially in the field of medical diagnostics. The development of an anatomical breast phantom confirmed the high sensitivity of the microwave radiometry method to temperature anomalies caused by tumour processes \cite{Polyakov2025}.
In the context of therapeutic applications, improved models have been proposed that outperform the classical Pennes equation in terms of accuracy in predicting temperature fields during laser exposure \cite{Sherief2024}. Analytical methods continue to play a crucial role in heat transfer studies, particularly in simplified cases. A generalised one-dimensional model of heat transfer between the core and the body surface made it possible to take into account the influence of temperature on perfusion and metabolic processes \cite{Kumar2021}. Another promising direction was the application of fractional calculus, which allowed us to obtain accurate solutions for spherical tissues and to analyse the influence of therapy parameters on thermal damage \cite{Hobiny2021}. Modern trends in modelling of thermal processes are characterised by the increasing complexity of the mathematical apparatus, consideration of anatomical features of tissues and close integration with diagnostic methods. Special attention is paid to verification of models on experimental data and development of physical phantoms, which allows increasing the accuracy of prediction of temperature distributions in clinically relevant problems \cite{Haemmerich2022, Akulova2024, Polyakov2025}. Continued improvement of computational algorithms and hardware opens up new opportunities for early diagnosis and optimisation of thermal treatments.

The majority of modern scientific works in this field are based on the Pennes biothermal balance equation \cite{Bardati2008, Avila2017, Figueiredo2019}. Despite some simplifications and approximate assumptions, this equation remains a reliable basis for modelling thermal processes in biological tissues, providing an adequate description of the temperature distribution inside the organism. In addition to the internal tissue temperature, a significant diagnostic parameter is the skin infrared (IR) temperature, which can indicate the presence of pathological changes. In addition, Lozano et al. \cite{Lozano2020} developed a computational model for analysing the thermal characteristics of breast cancer, based on high-resolution infrared imaging, three-dimensional reconstruction of the breast anatomy, and precise tumour localisation. By enabling a detailed analysis of the temperature profile of affected tissues, this model facilitates more accurate diagnosis and a personalised treatment approach for patients with a confirmed histological diagnosis.

Infrared thermography is a promising method of medical diagnostics, as it is not associated with exposure to ionising radiation and provides non-invasive examination of patients. This method is based on the registration and analysis of temperature distributions on the body surface, which makes it possible to detect abnormal heat signatures characteristic of various pathological processes. In recent years, significant progress in the development of diagnostic methods has been associated with the application of machine learning and neural network technologies, which improve the accuracy of pathology recognition and automate the analysis of thermographic images \cite{Goncalves2019, Torres2021, Manjunath2019}.
The study \cite{Figueiredo2018} proposes a methodology for identifying breast cancer using solely modelled thermal images. The experiments employed three-dimensional (3D) breast geometry derived from digital 3D scanning to provide high-fidelity modelling of thermal processes in tissues. In addition to the computer models, full-scale solid models containing an embedded heat source were used, allowing validation of the modelling at the physical level \cite{Igali2018}. One of the key challenges in thermal modelling of biological objects is the high computational complexity of such tasks. To efficiently compute thermal distributions, it is advisable to apply parallel computing methods, which can significantly reduce data processing time and increase the resolution of models \cite{Bousselham2018, Ewedafe2013}. In particular, the task of accurate localisation of the heat source inside biological tissues remains challenging and requires further development of computational algorithms and hardware solutions \cite{Yu2021}.

The aim of this work is to develop and study a computer model of thermal processes in biological tissues to analyse temperature distributions under various external conditions. The study considers the influence of the thermal conductivity coefficient, ambient temperature, and other parameters on the brightness temperature formed during measurements by microwave radiometry. The proposed model enables the evaluation of the influence of tissue thermal properties on the temperature distribution.

This work is organized as follows.
Section 2 outlines the materials and methods employed in the study, including the mathematical formulation of the heat transfer problem, the development of a multilayer geometrical model of biological tissues, and the numerical algorithms applied to solve the coupled biothermal–electrodynamic problem. Section 3 presents the results of computational experiments devoted to the analysis of the influence of environmental parameters and variations of thermophysical properties of tissues on the luminance temperature distribution. Section 4 discusses the results obtained and the main conclusions of the study, summarising the results of numerical simulations and outlining the directions for further work.

\section{Materials and methods}
\subsection{Mathematical model of brightness temperature of biological tissues during radiometric examinations}

Mathematical modelling of thermal processes in biological tissues, which are essentially inhomogeneous media, requires joint consideration of the problems of heat transfer and radiometric signal formation. The model is based on the non-stationary bio-thermal equation, which, for an isotropic medium, can be expressed as \cite{Haemmerich2022, Pennes1948}
\begin{equation}\label{eq:bioheat}
\rho(\mathbf{r}) c(\mathbf{r}) \frac{\partial T(\mathbf{r}, t)}{\partial t} = \nabla \cdot \left( k(\mathbf{r}) \nabla T(\mathbf{r}, t) \right) + q_m(\mathbf{r}, t) + q_b(\mathbf{r}, T) + q_{ext}(\mathbf{r}, t),
\end{equation}
where $T(\mathbf{r}, t)$ is the temperature of the tissue at the point $\mathbf{r}$ at time $t$, $\rho(\mathbf{r})$ is the density of the tissue, $c(\mathbf{r})$ is the specific heat capacity, $k(\mathbf{r})$ is the heat transfer coefficient, $q_m(\mathbf{r}, t)$ is the volumetric power of metabolic heat release, $q_b(\mathbf{r}, T) = \omega_b(\mathbf{r}) \rho_b c_b (T_a - T(\mathbf{r}, t))$ is the perfusion term describing heat exchange with blood flow ($\omega_b$ is the blood perfusion rate, $\rho_b, c_b$ is the density and heat capacity of the blood, $T_a$ is the arterial temperature), $q_{ext}(\mathbf{r}, t)$ is the power of external heat sources.
The convective heat transfer condition is set at the interface between the tissue and the environment
\begin{equation} \label{boundary}
	\lambda(\vec{r})\,\left(\vec{n}\cdot\nabla T(\vec{r},t)\right) = h\cdot (T_{air}-T(x, y, z, t)),
\end{equation}
where $\mathbf{n}$ is the vector of unit normal to the interface ``biological tissue~--~environment'', $h$ is the convective heat transfer coefficient (W/(m$^2 \cdot$K)), $T_{air}$ is the ambient air temperature.

The brightness temperature $T_B$ registered by the microwave radiometer at the frequency $\nu$ is determined by the spatial distribution of the thermodynamic temperature inside the irradiated tissue volume and its dielectric properties. For its calculation, an integral representation of the form is used \cite{Bardati2008, Vesnin2018}
\begin{equation}\label{eq:Tb_integral}
T_B(\nu) = \int\limits_{V_b} W(\mathbf{r}; \nu) T(\mathbf{r}), dV,
\end{equation}
denotes the tissue volume that contributes to the microwave signal received by the radiometer antenna. The weight function $W(\mathbf{r}; \nu)$ describes the contribution of the volume element $dV$ to the total radiometric signal and is defined by the equation
\begin{equation}\label{eq:weight_function}
W(\mathbf{r}; \nu) = \frac{P_d(\mathbf{r}; \nu)}{\int_{V_b} P_d(\mathbf{r}; \nu), dV},
\end{equation}
where $P_d(\mathbf{r}; \nu)$ is the electromagnetic power density dissipated in the tissue at the point $\mathbf{r}$ at the frequency $\nu$. The function $W(\mathbf{r}; \nu)$ satisfies the normalisation condition
$$
\int_{V_b} W\,dV = 1 .
$$

The dissipated power density is calculated by solving the electrodynamic problem of the electromagnetic field distribution in an inhomogeneous medium \cite{Fear2002, Polyakov2017}
\begin{equation}\label{eq:power_density}
P_d(\mathbf{r}; \nu) = \frac{1}{2}, \sigma(\mathbf{r}; \nu) \cdot |\vec{E}(\mathbf{r}; \nu)|^2,
\end{equation}
where $\sigma(\mathbf{r}; \nu)$ is the specific electrical conductivity of the tissue, $\vec{E}(\mathbf{r}; \nu)$ is the electric field strength vector. It is important to emphasise the strong dependence of dielectric properties of biological tissues ($\sigma$, $\varepsilon$) on the microwave radiation frequency $\nu$ \cite{Gabriel-1996}.

The spatial distribution of the electric field in the monochromatic limit is described by the Helmholtz equation
\begin{eqnarray}\label{eq-Gelmgoltz}
	\Delta\, \vec{E}+\frac{\omega^2}{c^2}  \varepsilon \,\vec{E} = - \nabla \left( \vec{E}\cdot \vec{\nabla} (\ln\,\varepsilon) \right),
\end{eqnarray}
where $\varepsilon(x,y,z;\nu)$ is the permittivity, $c$ is the speed of light in vacuum, $\omega=2\pi \nu$. 
The right-hand side of \mbox{equation \eqref{eq-Gelmgoltz}} clearly illustrates the effect of the inhomogeneity of the dielectric properties of biological tissue.

Thus, the complete mathematical model is a coupled boundary value problem combining the biothermia \mbox{equation (\ref{eq:bioheat})} with the boundary condition (\ref{boundary}) and the integral relation (\ref{eq:Tb_integral})-(\ref{eq:power_density}). 

\subsection{Model geometry}

\begin{figure}[h!]
\centering
\includegraphics[width=0.5\linewidth]{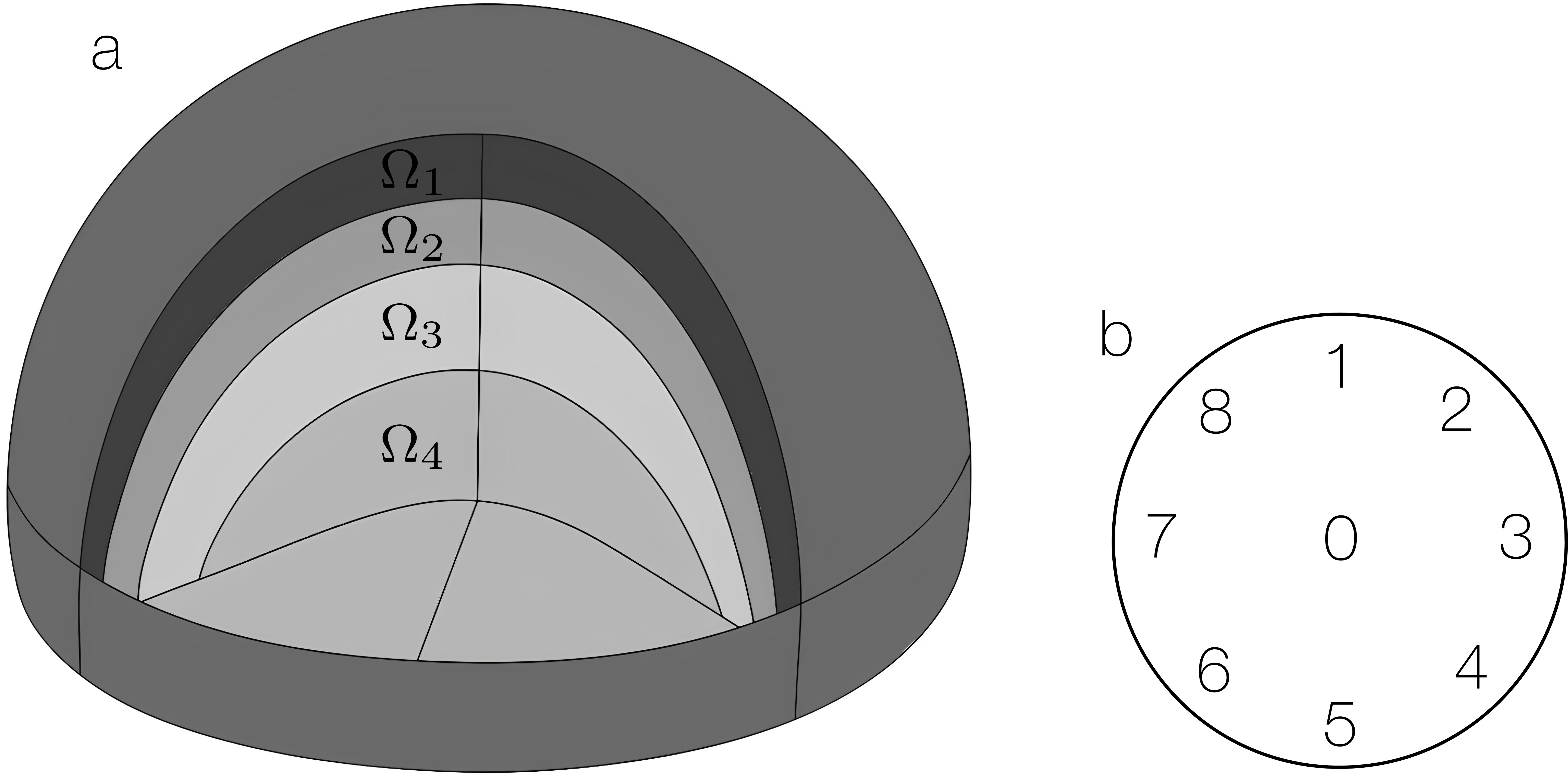}
\caption{Schematic representation of a multilayered geometric model of the breast: $\Omega_1$ is the skin, $\Omega_2$ is the subcutaneous adipose tissue, $\Omega_3$ is the glandular tissue, and $\Omega_4$ is the muscle tissue (a). The scheme of radiometric examinations of the breast (b).}
\label{fig:breast_geometry}
\end{figure}

A three-dimensional multilayer geometric model of the breast based on \cite{Lozano2020, Torres2021} anatomical data has been developed for mathematical modelling of thermal processes and microwave radiometry signal formation. The model consists of an axisymmetric hemispherical structure of radius $R$, comprising four isotropic, homogeneous layers with well-defined interfaces (Figure \ref{fig:breast_geometry}a). The choice of axisymmetric hemispherical geometry is dictated by considerations of reducing computational complexity while maintaining an adequate representation of heat transfer in the central part of the organ.

Mathematically, the model is described as a union of four non-intersecting subdomains, given by the equation $\Omega = \bigcup_{i=1}^ {4} \Omega_i$, where each subdomain is characterised by its own set of thermophysical $\{\rho_i, c_i, k_i\}$ and electrodynamic parameters $\{\varepsilon_i, \sigma_i\}$.

The geometric parameters of the model are specified in the spherical coordinate system $(r, \theta, \phi)$. The region $\Omega_1$, which represents the skin, is the outer spherical layer of thickness $d_1$, defined as $\Omega_1 = \{ (r, \theta, \phi) : R - d_1 \leq r \leq R,\ 0 \leq \theta \leq \pi/2,\ 0 \leq \phi < 2\pi \}$. Below the skin is the region $\Omega_2$, representing subcutaneous adipose tissue with thickness $d_2$, which is described by the relation $\Omega_2 = \{ (r, \theta, \phi) : R - d_1 - d_2 \leq r < R - d_1,\ 0 \leq \theta \leq \pi/2,\ 0 \leq \phi < 2\pi \}$. The main volume of the breast, the glandular tissue, forms the region $\Omega_3$ with a thickness of $d_3$, given as $\Omega_3 = \{ (r, \theta, \phi) : R_m \leq r < R - d_1 - d_2,\ 0 \leq \theta \leq \pi/2,\ 0 \leq \phi < 2\pi \}$, where $R_m = R - d_1 - d_2 - d_3$ is the radius of the inner boundary of the glandular tissue. The base of the model, represented as a spherical segment of thickness $d_4$ that models the pectoral muscle, forms the region $\Omega_4$, defined as $\Omega_4 = \{ (r, \theta, \phi) : R_m - d_4 \leq r < R_m,\ 0 \leq \theta \leq \pi/2,\ 0 \leq \phi < 2\pi \}$.

Boundary conditions are specified on two surfaces. On the outer surface $\partial \Omega_{ext}$, which is in contact with the air, the condition $-k_1 \nabla T \cdot \mathbf{n} = h (T - T_{\text{air}})$ is satisfied, where $h$ is the convective heat transfer coefficient and $T_{\text{air}}$ is the ambient air temperature. On the inner boundary $\partial \Omega_{base}$, which is in contact with the body, a constant temperature $T = T_{core}$ = 37 $^\circ \text{C}$ is maintained, corresponding to the core temperature of the body and the physiological condition of thermoregulation.

The model parameters and their typical values used in computational experiments are listed in Table \ref{tab:geometry_params}.

\vspace{2mm}
\noindent Table 1. Geometric parameters of a multilayer model of the breast.
\vspace{-4mm}
\begin{table}[h!]
\centering
\caption{Geometric parameters of a multilayer model of the breast}\label{tab:geometry_params}
\begin{tabular}{p{0.4\linewidth}c c}
\hline
Parameter & Symbol & Typical value \\
\hline
Radius of the breast & $R$ & 80 mm \\
Thickness of the skin layer & $d_1$ & 2.0 mm \\
Thickness of the fat tissue layer & $d_2$ & 10.0 mm \\
Thickness of the glandular tissue layer & $d_3$ & 50.0 mm \\
Muscle layer thickness & $d_4$ & 8.0 mm \\
Inner boundary radius & $R_m$ & 20.0 mm \\
\hline
\end{tabular}
\end{table}

The developed geometric model ensures anatomical accuracy and allows investigating the influence of variations in the thermophysical parameters of different tissues on the spatial distribution of temperature and the formation of a radiometric signal, which is key to optimising non-invasive thermodiagnostic methods \cite{Akulova2024, Polyakov2025}.

\subsection{Numerical methods and software}

Solving the coupled problem of heat transfer and electrodynamics requires joint consideration of the bioheat equation (\ref{eq:bioheat}) and the electromagnetic field equation (\ref{eq-Gelmgoltz}). For its numerical implementation, a specialised software package has been developed based on the finite element method in the spatial domain and the finite difference method for time integration. This approach allows for the correct consideration of the complex geometry of biological tissues and the significant heterogeneity of their thermophysical and electrodynamic properties.

The choice of the finite element method was motivated by its ability to handle complex geometries and to naturally incorporate boundary conditions. For time integration, an implicit Crank–Nicholson scheme of second-order accuracy is used, which ensures unconditional stability of the solution at reasonable computational cost.

\subsubsection{Discretisation of the heat transfer equation}

Spatial discretisation of the computational domain $\Omega$ is performed using tetrahedral finite elements that form the partition $\Omega = \bigcup_{e=1}^{N_e} \Omega^e$. Such elements allow complex shapes of biological structures to be adequately approximated. Within each element, the temperature field is approximated by linear functions of the form
\begin{equation}
T_h(\mathbf{x}, t) = \sum_{i=1}^{4} N_i(\mathbf{x}) T_i(t),
\end{equation}
where $N_i(\mathbf{x})$ are linear functions of the tetrahedral element, and $T_i(t)$ are the temperature values at the corresponding grid nodes.

The discrete form of the equation is based on a weak formulation of the problem, which relaxes the smoothness requirements of the solution and allows boundary conditions to be taken into account in a natural way. The original bioheat equation (\ref{eq:bioheat}) is multiplied by a test function $v(\mathbf{x})$ and integrated over the entire computational domain
\begin{equation}
\int_\Omega \rho c \frac{\partial T}{\partial t} v , d\Omega + \int_\Omega k \nabla T \cdot \nabla v , d\Omega = \int_\Omega Q v , d\Omega - \int_{\partial \Omega} k \frac{\partial T}{\partial n} v , dS,
\end{equation}
where $Q = q_m + q_b + q_{ext}$ is the total density of heat sources. Taking into account the boundary condition of convective heat transfer (\ref{boundary}) leads to the following weak formulation
\begin{equation}
\int_\Omega \rho c \frac{\partial T}{\partial t} v , d\Omega + \int_\Omega k \nabla T \cdot \nabla v , d\Omega + \int_{\partial \Omega} h T v , dS = \int_\Omega Q v , d\Omega + \int_{\partial \Omega} h T_{air} v , dS.
\end{equation}

After substituting the approximation of the temperature field and selecting the test functions, we obtain a system of ordinary differential equations
\begin{equation} \label{stiffnessmatrices}
[M] \frac{d{T}}{dt} + [K]{T} = {F},
\end{equation}
where $[M] = \bigcup_{e} \int_{\Omega^e} \rho c N_i N_j , d\Omega$ is the global mass matrix characterising the thermal inertia properties of the system;
$[K] = \bigcup_{e} \left( \int_{\Omega^e} k \nabla N_i \cdot \nabla N_j , d\Omega + \int_{\partial \Omega^e} h N_i N_j , dS \right)$ is the global stiffness matrix describing thermal conductivity and convective heat transfer;
and ${F} = \bigcup_{e} \left( \int_{\Omega^e} Q N_i , d\Omega + \int_{\partial \Omega^e} h T_{\text{air}} N_i , dS \right)$ is global load vector, including heat sources and boundary effects.

The system is integrated over time using the implicit Crank–Nicholson scheme, which provides second-order accuracy
\begin{equation} \label{matrixequation}
\left( \frac{[M]}{\Delta t} + \frac{[K]}{2} \right) {T}^{n+1} = \left( \frac{[M]}{\Delta t} - \frac{[K]}{2} \right) {T}^n + \frac{{F}^n + {F}^{n+1}}{2},
\end{equation}
where $\Delta t$ is the time step, and the superscripts correspond to different time layers.

\subsubsection{Calculation of the electrodynamic weight function}

To determine the weight function $W(r; \nu)$ required for calculating the brightness temperature, Helmholtz's equation is solved for the complex amplitude of the electric field at a given frequency $\nu$. This task is implemented using the finite difference method in the frequency domain with first-order absorbing boundary conditions. Based on the found distribution of the electric field, the dissipated power density and the corresponding weight function are calculated according to equations (\ref{eq:weight_function})–(\ref{eq:power_density}). Taking into account the spatial distribution of the dielectric properties of biological tissues at the operating frequency ensures correct modelling of the depth sensitivity of the radiometric system.

\subsubsection{Software implementation and verification of program code}

Software written in C++ was implemented to conduct computational experiments. The programme architecture was developed based on a modular principle and includes the following main components.

\begin{figure}[h!]
\centering
\includegraphics[width=0.98\linewidth]{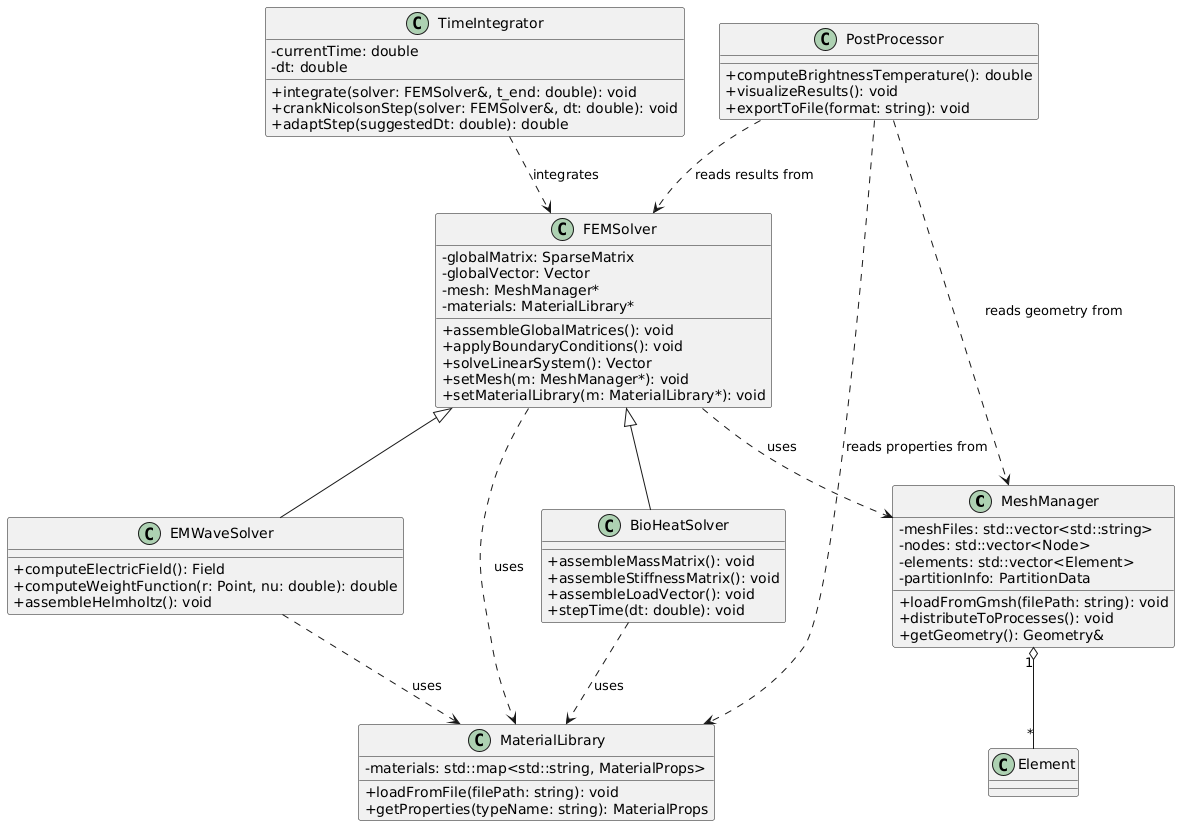}
\caption{Software class diagram for modelling the coupled problem of biothermics and electrodynamics}
\label{fig:class diagram}
\end{figure}

\begin{itemize}
\item The "MeshManager" class for working with computational mesh, responsible for loading tetrahedral meshes from GMSH files, distributing data between processes, and storing geometric information.
\item The "MaterialLibrary" material library containing thermophysical and electrodynamic parameters for various types of biological tissues.
\item The "FEMSolver" base class for solving problems using the finite element method, implementing procedures for assembling global matrices and vectors, applying boundary conditions, and solving systems of linear equations.
\item The "BioHeatSolver" is a subclass of "FEMSolver", specialised for solving the bioheat equation, implementing the assembly of mass matrices $[M]$ and stiffness matrices $[K]$, as well as load vectors $\{F\}$ according to equations \eqref{stiffnessmatrices}-\eqref{matrixequation}.
\item The "EMWaveSolver" class for solving electrodynamic problems, which calculates the electric field distribution and the weight function $\Omega(r; \nu)$ based on the Helmholtz equation.
\item The "TimeIntegrator" module for time discretisation, implementing the Crank-Nicholson scheme with adaptive time step selection.
\item The "PostProcessor" post-processing module, which calculates the brightness temperature using the formula \eqref{eq:Tb_integral} and visualises the results.
\item The "Element" is a base class representing a separate grid element into which a biological tissue model is divided.
\end{itemize}
The class diagram of the developed software is shown in Figure \ref{fig:class diagram}.

To solve systems of linear equations, the Cholesky decomposition for symmetric positive definite matrices and the multigrid method are used to accelerate convergence. Parallel computations are organised using OpenMP technology, which allows the computational load to be distributed between CPU cores.

\begin{figure}[h!]
\centering
\includegraphics[width=0.35\linewidth]{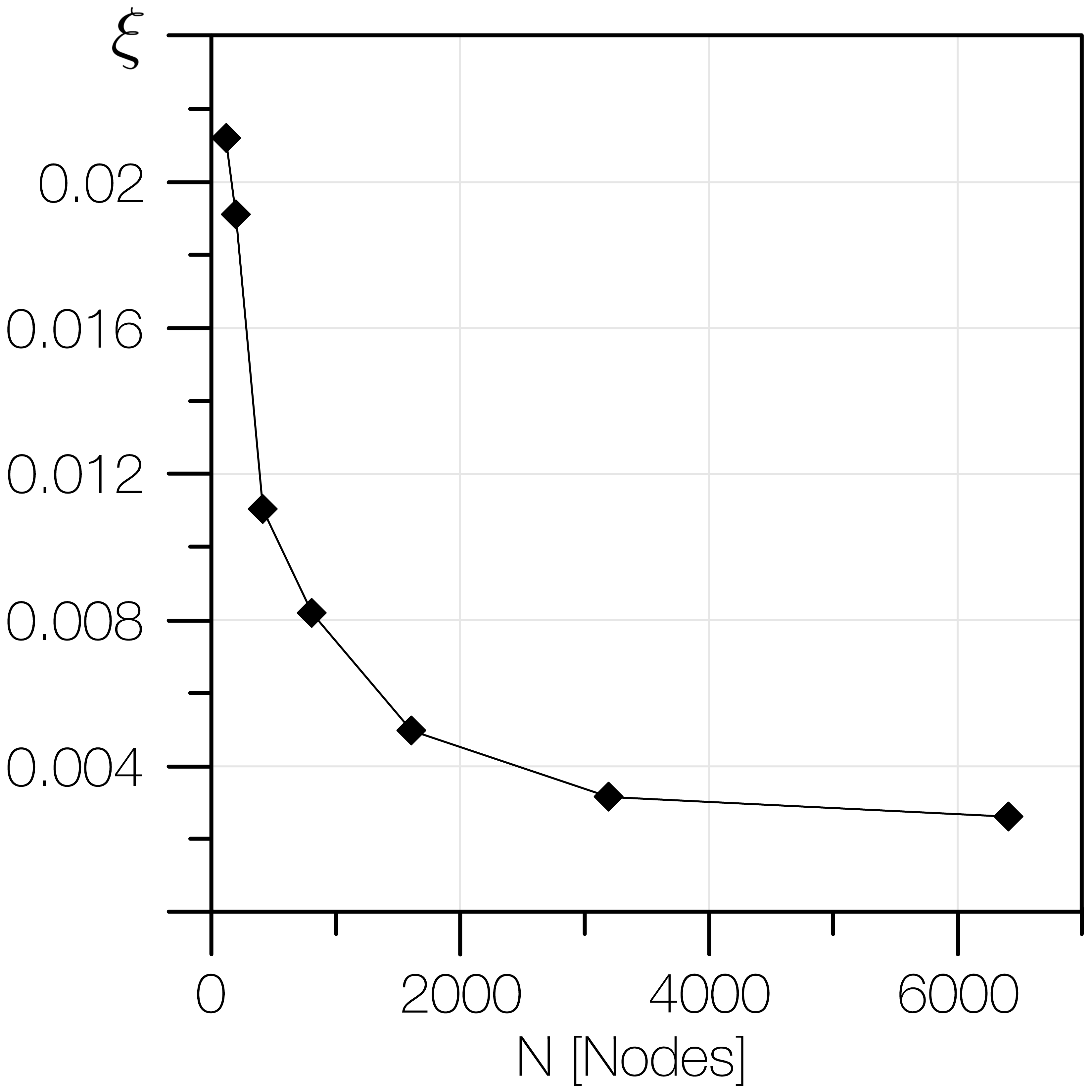}
\caption{Dependence of the maximum relative error of the numerical solution $\xi$ on the number of calculation cells $N$ (grid elements) during sequential refinement of the calculation grid}
\label{fig:convergence}
\end{figure}

The software package was verified. Individual modules were tested on analytical solutions for heat conduction problems in a homogeneous sphere and electromagnetic wave propagation in a layered medium. The convergence of mesh solutions was also checked during sequential refinement of the calculation mesh. The dependence of the relative error on the number of calculation cells $N$ was calculated \mbox{using the ratio}
\begin{equation}\label{slae1}
	\xi=\frac{1}{N}\sum_{i=1}^N \frac{|T_{i}^{num}-T_{i}^{an}|}{T_{i}^{an}},
\end{equation}
where $T_{i}^{num}$ is the numerical solution at node $i$, and $T_{i}^{an}$ is the analytical solution at node $i$. Figure \ref{fig:convergence} shows the convergence of the numerical solution as $N$ increases, where $\xi_{max}\approx$ 0.021 for $N$=100 and $\xi_{min}\approx$ 0.004 for $N$=6500.

The verification results showed that the error in calculating temperature fields does not exceed 0.1 $^\circ$C, and the order of convergence of the method corresponds to the theoretical value for the linear elements used and the second-order accuracy scheme. Considering that the error of the microwave radiometry method is about 0.2 $^\circ$C, it can be concluded that the numerical methods used provide the necessary accuracy for solving the diagnostic tasks.

\section{Results of computational experiments}

This section presents the results of a numerical analysis of the influence of key parameters of the developed biothermal model on the distribution of brightness temperature. The study was conducted using a three-dimensional multilayer model of the breast, whose anatomical structure is characterised by heterogeneity and variability in its thermophysical properties. The main focus is on assessing the sensitivity of brightness temperature to changes in environmental parameters, as well as to variations in the thermal conductivity coefficient of biological tissues.

\subsection{The influence of environmental parameters on the distribution of brightness temperature}

An important aspect affecting the accuracy of microwave radiometry is the external conditions under which the survey is conducted. This experiment investigates the influence of two key environmental parameters: air temperature ($T_{air}$), and convective heat transfer coefficient ($h$) on the distribution of brightness temperature ($T_B$) in a multilayer model of the breast.

To assess this influence, a series of computational experiments were conducted, varying $T_{air}$ in the range from 18 to 26 $^\circ$C and $h$. $T_B$ was recorded at several points in the model (see Figure \ref{fig:breast_geometry}b).

\begin{figure}[h!]
\begin{center}
\includegraphics[height=6cm]{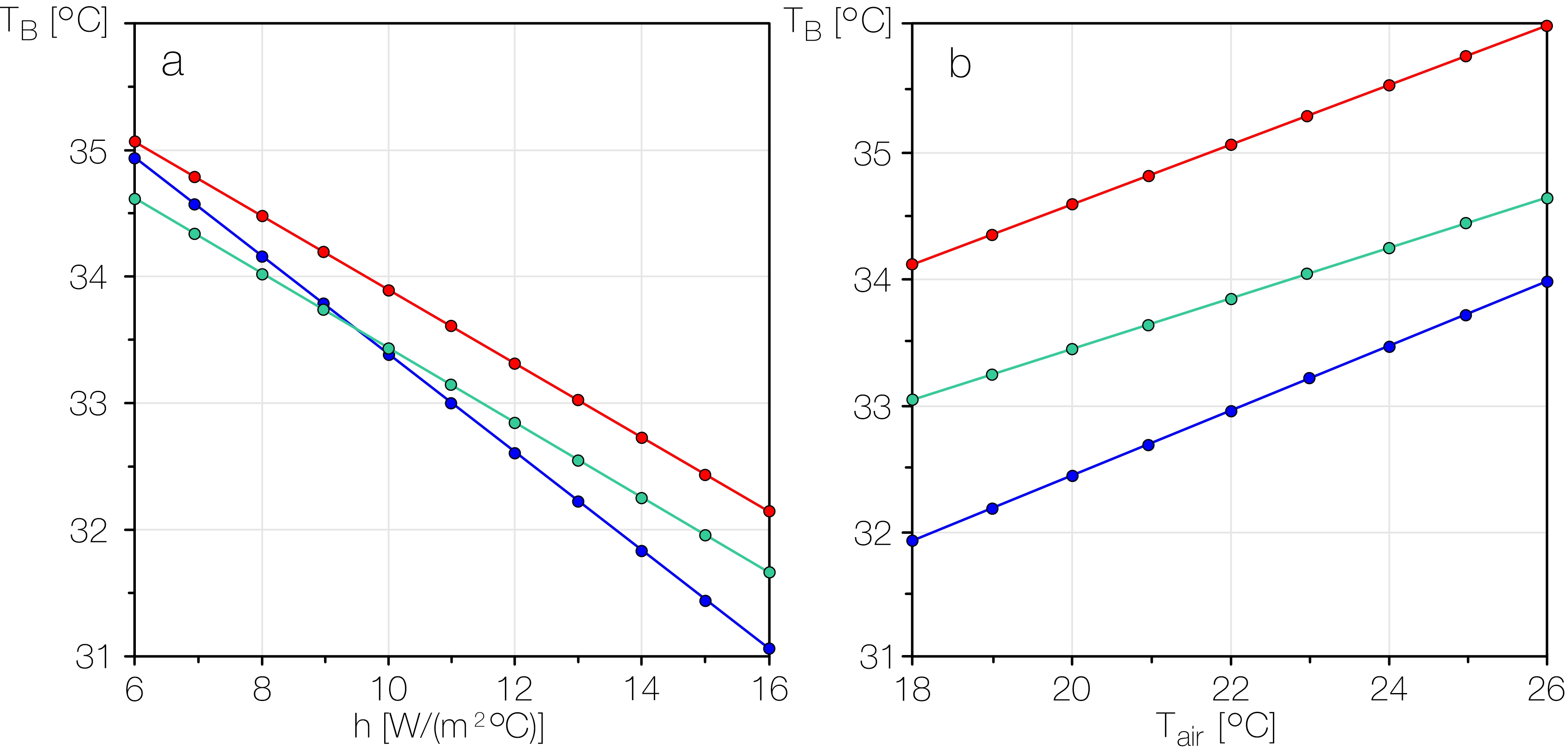}
\caption{Dependence of brightness temperature $T_B$ on heat transfer coefficient $h$ (a); ambient temperature $T_{air}$ (b): the red line corresponds to point "0" on the breast, the blue line corresponds to point "3", and the green line corresponds to point "7".}
\label{fig1}
\end{center}
\end{figure}

The simulation results shown in Figure 4 demonstrate a pronounced linear dependence of the brightness temperature on both parameters. Figure 4a shows the dependence of $T_B$ on the heat transfer coefficient $h$ at a fixed air temperature. An increase in brightness temperature is observed with an increase in $h$, which is explained by the intensification of convective cooling of the surface and, as a result, a change in the temperature gradient into the tissue. The dependence is observed for all points of the model. Figure 4b shows the dependence of $T_B$ on the ambient temperature $T_{air}$ at a constant heat transfer coefficient. An increase in $T_{air}$ leads to a proportional increase in $T_B$ at all measurement points. Quantitative analysis reveals that an 8 $^\circ$C variation in $T_{air}$ leads to an approximately 2 $^\circ$C change in $T_B$, highlighting the importance of strict room temperature control during diagnostic procedures.

\begin{figure}[h!]
\begin{center}
\includegraphics[height=7cm]{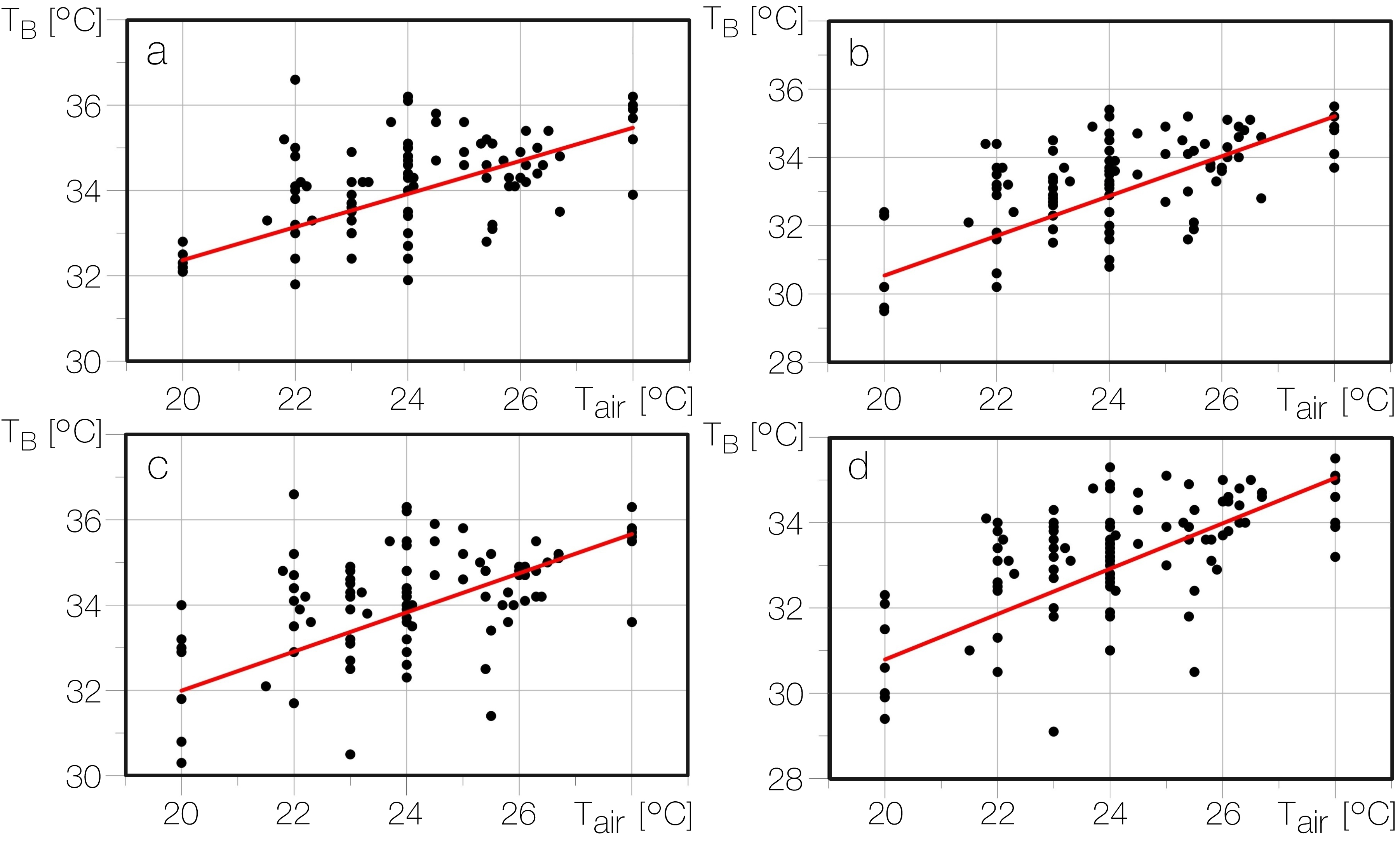}
\caption{Dependence of brightness $T_B$ and infrared $T_{IR}$ temperatures on ambient temperature $T_{air}$ in field experiments for point "0": right breast (a-b), left breast (c-d).}
\label{fig5}
\end{center}
\end{figure}

To validate the computational model, the results of numerical calculations were compared with data from field experiments. Figure \ref{fig5} shows the dependence of brightness ($T_B$) and infrared ($T_{IR}$) temperatures on air temperature ($T_{air}$), measured for the right (a, b) and left (c, d) breast at point "0". A comparison of the graphs shows qualitative and quantitative agreement between the dependencies obtained during numerical modelling and \textit{in vivo} measurements. The observed parallelism in the behaviour of $T_B$ and $T_{IR}$ in response to changes in external conditions confirms the adequacy of the developed mathematical model and the correctness of the boundary conditions.

Thus, the established dependence of brightness temperature on external parameters dictates the need for strict control and standardisation of environmental conditions in the clinical diagnostic process to ensure the reproducibility and reliability of microwave radiometry results.

\subsection{The influence of variations in the thermal conductivity coefficient on the brightness temperature}

The thermal conductivity of biological tissues shows considerable variability due to their structural and physicochemical characteristics. Tissues with high water content, such as muscle and glandular tissue, are characterised by thermal conductivity coefficients $k$ close to that of water. At the same time, tissues with low density and high lipid content, in particular subcutaneous adipose tissue and the stratum corneum of the skin, have significantly lower $k$ values, occupying an intermediate position between aqueous environments and air. These differences form a complex thermal insulation system that regulates heat exchange between internal organs and the environment.

\begin{figure}[h!]
\begin{center}
\includegraphics[height=6cm]{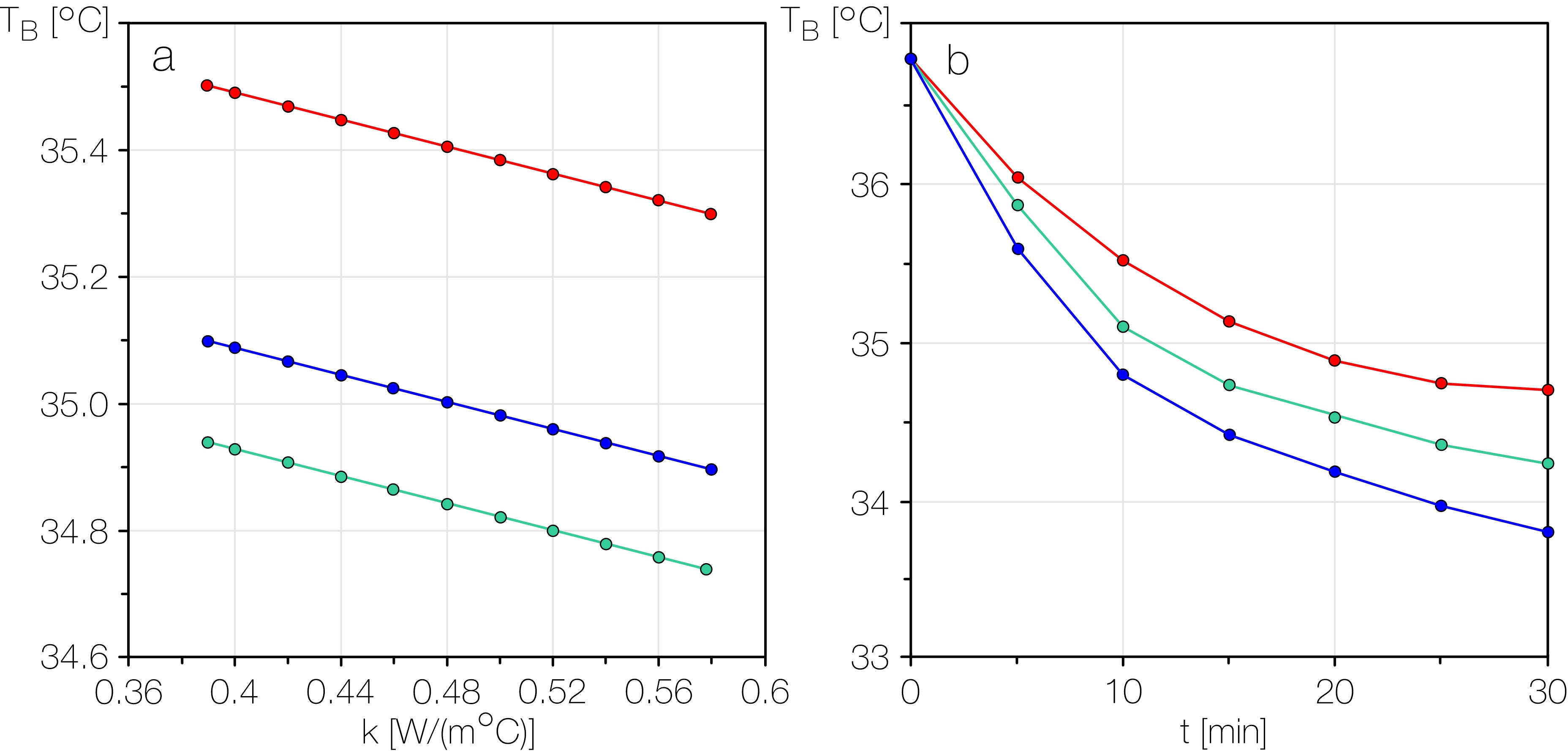}
\caption{Dependence of brightness temperature $T_B$ on thermal conductivity coefficient $k$ (a); temperature distribution at point "0" for different values of the average thermal conductivity coefficient (b): the red line corresponds to $k$ = 0.4 W/(m$\cdot^{\circ}$C), the green line corresponds to $k$ = 0.5 W/(m$\cdot^{\circ}$C), the blue line is $k$ = 0.6 W/(m$\cdot^{\circ}$C) (without taking into account the nonlinear nature of $q_{m}$).}
\label{fig2}
\end{center}
\end{figure}

Muscle tissue, which has relatively high thermal conductivity and a developed vascular network, provides effective heat transfer from deep structures to the periphery, performing a thermoregulatory function. In contrast, the skin and fat layers create a thermal barrier that minimises heat loss to the environment. This mechanism maintains thermal homeostasis and explains the observed temperature distribution between the body core and its surface. The results of parametric analysis presented in Figure \ref{fig2} reveal quantitative patterns of the influence of thermal conductivity on the formation of the radiometric signal.

Figure \ref{fig2}a shows the linear dependence of the brightness temperature $T_B$ on the thermal conductivity coefficient $k$, obtained for the measurement point "0". An increase in $k$ by 0.2 W/(m$\cdot^{\circ}$C) leads to an increase in $T_B$ by approximately 1.5 $^\circ$C, which indicates the high sensitivity of the method to changes in the thermophysical parameters of tissues. Figure \ref{fig2}b shows the temperature profiles calculated for different values of $k$. Analysis of the distributions shows that as thermal conductivity increases, the temperature field becomes significantly more uniform: temperature gradients decrease and heat flow from deep layers to the surface intensifies. The results obtained are in qualitative agreement with the data presented in \cite{Luitel} and confirm the importance of accurately determining the thermophysical characteristics of biological tissues for the interpretation of microwave radiometry data.

\section{Discussion and Conclusion}

The results of numerical modelling demonstrate the high sensitivity of brightness temperature, recorded by microwave radiometry, to changes in environmental parameters and the thermophysical properties of biological tissues. The established linear dependence of $T_B$ on air temperature $T_{air}$ and convective heat transfer coefficient $h$ is consistent with data obtained in other studies \cite{Bardati2008, Sedankin2018}. In particular, it is shown that a change in $T_{\text{air}}$ by 8$^\circ$C leads to a significant shift in $T_B$ by approximately 2$^\circ$C, which exceeds the typical error of the MWR method (about 0.2$^\circ$C) and confirms the need for strict thermostatting of the diagnostic room.

The identified dependence of brightness temperature on the thermal conductivity coefficient $k$ also has important practical significance. The results obtained correlate with the data from \cite{Haemmerich2022}, which emphasizes the significant variability of the thermophysical parameters of biological tissues and their influence on temperature distributions. The linear nature of the dependence $T_B(k)$ observed in our study indicates that even natural variations in tissue thermal conductivity can contribute significantly to the radiometric signal, which must be taken into account when diagnosing pathologies associated with changes in the histological structure of tissues, such as malignant \mbox{neoplasms \cite{Lozano2020}.}

The developed multilayer model, which takes into account the anatomical features of the breast structure, provides a more accurate description of thermal processes compared to homogeneous models \cite{Figueiredo2019}. The joint solution of biothermics and electrodynamics equations allows for the correct determination of the weight function $\Omega(r; \nu)$, which is a significant advantage in modelling deep temperature. Verification of the model using analytical solutions and data from field experiments confirmed its adequacy and accuracy, which is consistent with the approaches presented in the works \cite{Vesnin2018, Polyakov2025}.

One limitation of the present model is the assumption of constant metabolic heat production $q_{m}$. As shown in studies \cite{Marin2021, Sherief2024}, taking into account the nonlinear dependence of metabolic and perfusion parameters on temperature can improve the accuracy of modelling, especially in areas with impaired vascularization. In addition, the model does not take into account possible spatial variations in blood perfusion velocity, which can be significant when modelling areas with tumor formations \cite{Gautherie1982}.

Promising areas for further research include integrating the developed model with machine learning methods for automatic interpretation of radiometric data \cite{Levshinskii2020}, as well as accounting for relaxation effects of heat transfer based on models of nonlocal and fractional transport \cite{Ragab2021, Hobiny2021}. The developed software package can be adapted for modelling other organs and pathological conditions, as well as for planning and optimizing thermal treatment methods.

This study yielded the following main results:

1) A comprehensive mathematical model of thermal processes in multilayer biological tissues has been developed, combining the Penna bioheat equation with an electrodynamic model of radiometric signal formation. The model takes into account the anatomical features of the breast structure and the variability of tissue thermophysical parameters.
    
2) A significant influence of environmental parameters on brightness temperature has been established. It has been shown that a change in air temperature of 8 °C causes a shift in brightness temperature of approximately 2 °C, which exceeds the error margin of the microwave radiometry method and confirms the need for strict control of the conditions under which diagnostic procedures are performed.
    
3) A linear dependence of brightness temperature on the thermal conductivity coefficient of biological tissues has been identified. An increase in $k$ by 0.2 W/(m·$^\circ$C) leads to a decrease in $T_B$ by approximately 1.5 $^\circ$C, which demonstrates the high sensitivity of the method to variations in the thermophysical characteristics of tissues.
    
A dedicated software package based on the finite element method was developed and validated against analytical solutions and experimental data. The error in temperature field calculations does not exceed 0.1 $^\circ$C, which provides sufficient accuracy for diagnostic tasks.

The results obtained provide a basis for further improvement of computer modelling methods in medical diagnostics and can be used in the development of clinical protocols for microwave radiometry.

\textbf{Funding:} This work supported by the Russian Science Foundation (grant № 25-21-00330, \href{https://rscf.ru/project/25-21-00330/}{https://rscf.ru/project/25-21-00330/}).

\end{document}